\documentclass[letterpaper]{article} 
\usepackage[]{aaai24}  
\usepackage{times}  
\usepackage{helvet}  
\usepackage{courier}  
\usepackage[hyphens]{url}  
\usepackage{graphicx} 
\usepackage{natbib}  
\usepackage{caption} 
\usepackage{algorithm}
\usepackage{algorithmic}
\usepackage{amssymb}

\usepackage{pgfplots}
\usepgfplotslibrary{groupplots}
\usepgfplotslibrary{fillbetween}
\usepackage{filecontents}
\usepackage[T1]{fontenc}
\usetikzlibrary{math}
\tikzmath
{
  function symlog(\x,\a){
    \yLarge = ((\x>\a) - (\x<-\a)) * (ln(max(abs(\x/\a),1)) + 1);
    \ySmall = (\x >= -\a) * (\x <= \a) * \x / \a ;
    return \yLarge + \ySmall ;
  };
  function symexp(\y,\a){
    \xLarge = ((\y>1) - (\y<-1)) * \a * exp(abs(\y) - 1) ;
    \xSmall = (\y>=-1) * (\y<=1) * \a * \y ;
    return \xLarge + \xSmall ;
  };
}

\pgfplotsset{every axis/.append style={
                    xlabel={$x$},          
                    ylabel={$y$},          
                    label style={font=\sffamily},
                    tick label style={font=\sffamily\scriptsize},
                    xticklabel style = {font=\sffamily\scriptsize},
                    title style = {font=\normalsize\sffamily},
                    ylabel near ticks,
                    y label style={font=\sffamily\scriptsize},
                    xlabel near ticks,
                    x label style={font=\sffamily\scriptsize},
                    legend cell align={left},
                    legend style={draw=none, font=\sffamily\scriptsize},
                    },
                    legend image code/.code={
                    \draw[mark repeat=2,mark phase=2]
                        plot coordinates {
                        (0cm,0cm)
                        (0.15cm,0cm)        
                        (0.3cm,0cm)         
                        };%
                    }
                    }
\pgfplotsset{compat=newest}

\usepackage{newfloat}
\usepackage{listings}
\DeclareCaptionStyle{ruled}{labelfont=normalfont,labelsep=colon,strut=off} 
\floatstyle{ruled}
\newfloat{listing}{tb}{lst}{}
\floatname{listing}{Listing}
\title{Reputation Transfer in the Twitter Diaspora}

\author {
    Kristina Radivojevic,
    DJ Adams,
    Griffin Laszlo,
    Felixander Kery,
    Tim Weninger
}
\affiliations {
    Department of Computer Science and Engineering \\
    University of Notre Dame\\
    \{kradivo2, dadams22, glaszlo, fkery, tweninger\}@nd.edu
}

\usepackage{bibentry}

\begin{document}

\maketitle

\begin{abstract}
Social media platforms have witnessed a dynamic landscape of user migration in recent years, fueled by changes in ownership, policy, and user preferences. This paper explores the phenomenon of user migration from established platforms like X/Twitter to emerging alternatives such as Threads, Mastodon, and Truth Social. Leveraging a large dataset from Twitter, we investigate the extent of user departure from X/Twitter and the destinations they migrate to. Additionally, we examine whether a user's reputation on one platform correlates with their reputation on another, shedding light on the transferability of digital reputation across social media ecosystems. Overall, we find that users with a large following on X/Twitter are more likely to migrate to another platform; and that their reputation on X/Twitter is highly correlated with reputations on Threads, but not Mastodon or Truth Social. 
\end{abstract}

\section{Introduction}

\begin{figure}
    \centering
    \includegraphics[width=.9\linewidth]{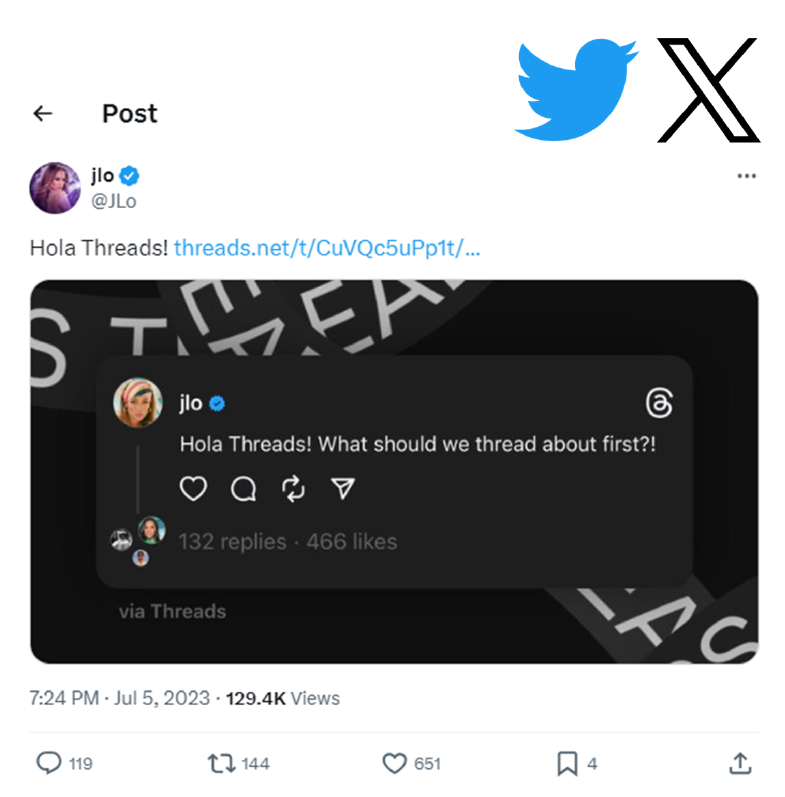}
    \caption{Example of a user migrating from X/Twitter to Threads. In the present work we find that users with large following are likely to migrate from X/Twitter to another platform. We also find that their reputations (in the number of followers) tends to transfer to Threads, but not to Mastodon or Truth Social.}
    \vspace{-.5cm}
    \label{fig:lopez}
\end{figure}

The evolution of online social platforms has been marked by a cyclical pattern of rise and fall, with users seeking new spaces for expression and engagement. From the early days of MySpace and Facebook, from Digg to Reddit, and from Vine to TikTok, the landscape has witnessed numerous transitions as users migrate in search of fresh experiences and communities. More recently, the once-dominant X/Twitter platform has faced a notable exodus of users to emerging platforms like Mastodon, Truth Social, and Threads. Consider the example from Fig.~\ref{fig:lopez}, where the musician and actor Jennifer Lopez, is one of many who have migrated from X/Twitter for alternate platforms recently. This shift raises intriguing questions about the transferability of users' reputations as they traverse from one platform to another. How does one's influence and standing on a particular platform carry over to new digital spaces? Exploring these dynamics offers valuable insights into the evolving nature of online discourse, interaction, and community dynamics.

Social media platforms are widely recognized as valuable tools for fostering democratic discourse by facilitating participation, disseminating political knowledge, and fostering trust among users~\cite{fujiwara2023effect,ding2023electoral}. However, they also give rise to challenges such as polarization, populism, and the proliferation of echo chambers. The ease of content creation on these platforms has exacerbated concerns regarding the reliability and trustworthiness of the information shared.


To ascertain the whether social media use is reflected in real world processes, researchers often analyze feedback mechanisms, such as digital reputation, within large-scale networks. Digital reputation, often reflected in metrics like follower counts and engagement levels, can shape individuals' influence and credibility within online communities. Consequently, some users may strategically manipulate their digital reputations to enhance their online influence, often through tactics like purchasing followers. However, these reputational dynamics may not seamlessly translate across platforms, leading to disparities in users' perceived trustworthiness.

As a result, individuals may enjoy high levels of credibility on one platform while having minimal presence on another, raising questions about the portability and reliability of digital reputations across different online ecosystems.

It is essential to distinguish between \textit{reputation} and \textit{influence}. User influence refers to how others perceive a user's presence and impact on a platform. Although trust, credibility, and influence may stem from reputation, they are not inherently linked. An individual's reputation can be shaped by various factors, including their values and the persona they project online. However, influence does not necessarily accompany reputation. In the present work, we focus on reputation. Rather than comparing the identities of followers migrating between platforms, we focus on the scale of followers gained on the new platform. We define reputation transfer as the efficacy of a user's reputation from one platform in establishing a reputation on another platform.

Previous studies have delved into the underlying motivations driving platform migration, exploring the factors that pull users towards new platforms while simultaneously pushing them away from existing ones~\cite{zengyan2009cyber,jeong2023exploring}. Additionally, research has investigated the typical behaviors exhibited by migrants during such transitions~\cite{jeong2023exploring}, as well as the ramifications of deplatforming – the removal of users from social media platforms for violating community guidelines or platform rules~\cite{rogers2020deplatforming}.

The acquisition of Twitter by Elon Musk in October 2022 sparked widespread attention and debate. Musk's outspoken advocacy for free speech absolutism has garnered admiration from some users, while others have voiced concerns about the potential exacerbation of misinformation on the platform. This polarizing event prompted a significant number of users to seek refuge on alternative platforms. Conversely, the act of deplatforming certain individuals from mainstream platforms has also triggered strong reactions from their followers. In response, these banned users and their supporters have gravitated towards platforms that align more closely with their political ideologies, often bringing their followers along with them.

In light of recent developments, including the ownership change at X/Twitter and the emergence of competing platforms such as Threads, Mastodon, and Truth Social, several key questions have emerged:

\textbf{RQ1:} How many users from X/Twitter have migrated away from the platform, and what alternative platforms have they chosen?

\textbf{RQ2:} How does reputation, in the form of followers and interactions, transfer when users migrate between social media platforms?

To address these questions, we use a comprehensive dataset gathered from X/Twitter, enabling us to identify users who have transitioned to Threads, Mastodon, and Truth Social. We track the migration patterns of these users and analyze the transfer of their reputation across platforms subsequent to their migration. Our findings suggest that (\textbf{RQ1}) users with a substantial following on X/Twitter are more inclined to migrate to alternative platforms. Furthermore, we observe that (\textbf{RQ2}) their reputations, as measured by the number of followers, tend to transfer to Threads, but not to Mastodon or Truth Social.

\section{Related Work}

Our exploration of social media migration and reputation is structured around two fundamental aspects. First, we delve into the relationship between reputation and influence within social media platforms. Understanding how these factors operate and intersect is essential for comprehending the transferability of user reputation across various platforms. Second, we examine the broader landscape of migrations and deplatforming events, aiming to uncover their implications for user reputation management. 

\subsection{Reputation and Influence}

Reputation and influence serve multifaceted roles on the internet, particularly within social media platforms, where they are utilized for various purposes such as political engagement \cite{vaccari2015political}, financial endeavors \cite{tsikerdekisonline}, and spreading rumors \cite{oh2013community}. An individual's reputation, representing how they are perceived by the public, is pivotal in fostering trust and, consequently, wielding influence. According to social impact theory, influence denotes the manifestation of social impact resulting in behavioral change following social interactions, whereas reputation signifies the magnitude of impact, which escalates with the source's strength \cite{latane1981social}.

Reputation and influence serve as primary indicators of leadership within the realm of social media, as illustrated by figures like Donald Trump, who transitioned to Truth Social after being banned from Twitter; Elon Musk, renowned for his Cryptocurrency endorsements~\cite{hamurcu2022can}, and Pope Francis, who has actively engaged with the public through social media platforms~\cite{cardoso2019media}. The actions of such leaders wield considerable influence over social media discourse and can trigger migrations between platforms. Consequently, scholars have extensively explored methods to identify these leaders and their followers~\cite{shafiq2013identifying}, seeking to forecast migration patterns that may impact reputation dynamics. 

\subsection{Migrations, Deplatforming, and Its Effect on Reputation}

The phenomenon of migration between social media platforms has garnered significant attention from researchers, tracing back to notable shifts from MySpace to Facebook~\cite{robards2012leaving}, Digg to Reddit~\cite{lerman2006social}, and from Facebook to Instagram~\cite{chan2020facebook}, among others. Central to the study of these migrations is the push-pull theory~\cite{zengyan2009cyber}, which posits that various factors influence individuals' decisions to transition from one platform to another~\cite{gerhart2019social}. These factors may include platform design, toxicity levels, moderation policies, and the presence of one's social circle or community~\cite{fiesler2020moving}.

Platform acquisitions also play a pivotal role in migration dynamics, as exemplified by Elon Musk's acquisition of X/Twitter, which spurred research by \citet{zia2023flocking}. Their analysis of Twitter migration to Mastodon, a decentralized social media platform, revealed that 2.26\% of users completely left Twitter following the acquisition, and users who remained active on both platforms exhibited distinct posting behaviors. Similarly, \citet{matthews2023politicians} investigated the migration of Twitter users to Parler, highlighting its significant impact on political polarization despite its limited user base. Prior research by \citet{kumar2011understanding} laid the groundwork for identifying migration patterns, paving the way for subsequent studies such as that by \citet{jeong2023exploring} to further identify migration dynamics.

When individuals or groups violate platform rules, social media companies may resort to ``deplatforming'' as a means of enforcement~\cite{chandrasekharan2017you, jhaver2021evaluating,thomas2021behavior}. Notably, in 2019 several influential figures faced removal from Facebook and Instagram due to involvement in organized hate and violence~\cite{MIT, facebook}. Subsequently, some of these individuals reported significant impacts on their follower count, reputation, and overall influence \cite{milo}. In response to such bans, affected users often migrate to platforms aligned with their political ideologies, exemplified by movements to platforms like Gab, known for its commitment to free speech~\cite{kalmar2018twitter}, or Telegram~\cite{rogers2020deplatforming, bryanov2021other}.


\begin{figure}[t]
    \centering
    \includegraphics[width=\linewidth]{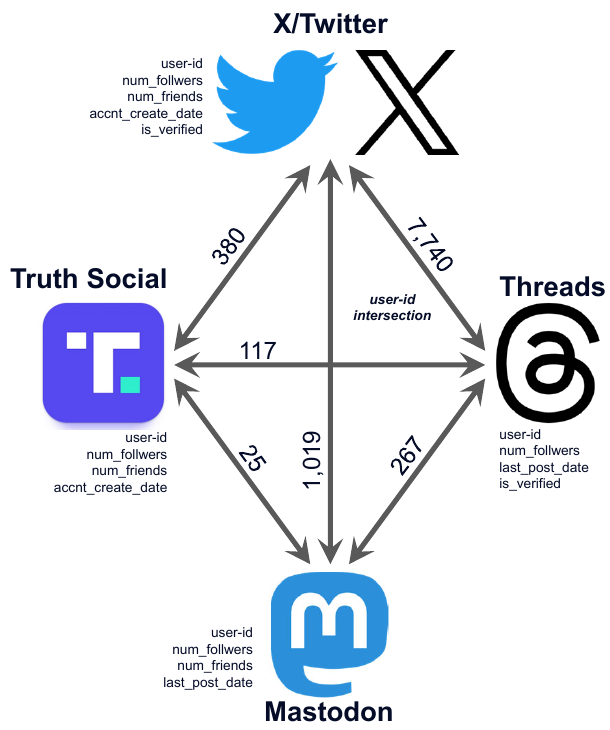}
    \caption{Dataset, including sizes of intersecting username sets from the refined collection of (N=117,919) X/Twitter users.}
    \label{fig:disapora}
\end{figure}

\section{Collecting Migrations}

In this section, we describe our methodology for collecting data on user migrations across multiple social media platforms and analyzing reputation transfer among them. Our initial dataset was created by merging datasets from X/Twitter, sourced from open source datasets \citet{mckelvey2017aligning} and \citet{twitterData}. From the initial pool of 224,296 user-ids in this combined dataset, we specifically identified user-ids that were found (\textit{i.e.}, were an identical match) in either Threads, Mastodon, or Truth Social. This filtering process resulted in a final dataset comprising 117,919 X/Twitter user-ids. This dataset, including the intersections of user-ids between the various platforms and the specific data collected, is illustrated in Fig.~\ref{fig:disapora}.

\paragraph{X/Twitter} For our analysis, we needed to supplement the initial dataset obtained from X/Twitter, which was created from datasets dating back to 2015 and 2017. To ensure our data was current, we conducted additional crawling of Twitter users who were still present (although perhaps not active) on the platform at the time of our analysis. From the refined X/Twitter dataset (N=117,919), we randomly selected a subset of 10,000 users. Using the ScrapFly API \cite{scrapfly}, we crawled X/Twitter and identified 6,340 accounts that were still present on X/Twitter as of February 2024. During this collection process, we also identified the user's follower count, friends count, account creation date, and verification status.

\paragraph{Threads} From the refined X/Twitter dataset (N=117,919) we queried Threads for user information. This process yielded 7,740 Threads accounts that shared identical usernames with X/Twitter. We also collected user information including follower count, verification status, and last post date.

\paragraph{Mastodon} The same process was used to collect data from Mastodon. This effort identified 1,019 accounts with matching usernames on both platforms. We also collected user information including follower count, friends count, and the last post date of each user.

\paragraph{Truth Social} 
For the Truth Social platform, rather than scraping the site directly, we used an existing dataset \cite{gerard2023truth} containing information on 454,000 users. From this dataset, we identified 380 users with the same usernames as those in the refined Twitter dataset. For each user, we gathered data on follower count, friends count, and account creation date.

\section{Migration Analysis}
\subsection{Methodology}

To initiate our analysis, we began by compiling a union dataset consisting of users from the Twitter dataset that had identical usernames across Mastodon, Threads, and Truth Social platforms, as described in the preceding section. From the original dataset with 114,000 usernames, we identified 9,544 Twitter accounts that were also present on at least one of the alternative platforms.

We rescraped the Twitter data, commencing on April 5, 2024, and continuing until April 9, 2024, utilizing the union list of usernames. Through this process, we identified 8,347 users that still had accounts on the Twitter platform---although these accounts may or may not have been active. These 8,347 accounts constitute the focal point of our analysis, representing migrated users who have transitioned from Twitter to alternative platforms.

To establish a control group, we selected 10,000 users from the original 117,919 X/Twitter dataset. Subsequently, we conducted a thorough rescraping of this control set and identified 6,469 users who still had accounts on Twitter, but did not have any accounts on the alternative platforms.

In summary, the treatment group contained 8,347 Twitter accounts that also had an account of the same username on at least one other platform; the control group contained 6,469 Twitter accounts that did not have an account (with the same username) on one of the other platforms.

The process of matching users between platforms introduces a notable threat to the validity of this methodology. Although this serves as a practical method for identifying users across multiple platforms, it inherently assumes that users maintain consistent usernames across different social media accounts. However, this assumption may not always hold true, as individuals might opt for different usernames or aliases on different platforms for various reasons, such as privacy concerns, branding strategies, or simply due to availability. Consequently, users with disparate usernames across platforms may not be accurately captured in the dataset. However, any bias inherent in exact username matching is consistent between the treatment and control sample methodologies; therefore although our results are certainly not complete, we do not expect severe limits in our statistical analysis.

\begin{table}[t]
\begin{center}
\caption{Logistic regression analysis of whether users migrated to another platform (encoded as 1) or not (encoded as 0). Users with a larger number of followers is are significantly more likely to migrate to another platform. The following count and tenure on X/Twitter (in months) is also weakly (but statistically significantly) correlated with migration as well.}
\vspace{-.1cm}
\small
\label{tab:logits}
\begin{tabular}{rrrrrl}
\hline
\multicolumn{3}{l}{Log-Likelihood:	-7,675.50} & \multicolumn{3}{r}{$N$: 13,280} \\
& \textbf{$\beta$} & \textbf{Std. Err} & \textbf{t-value} & \textbf{Pr($>$$|$t$|$)} &   \\ 
\hline
Const. & -2.5271 & 0.099 & -25.577 & 0.000 & *** \\ 
Followers & 0.2030 & 0.010 & 21.140 & 0.000 & *** \\ 
Following & 0.0124 & 0.001 & 18.214 & 0.000 & *** \\ 
Months & -0.0409 & 0.014 & -2.882 & 0.004 & ** \\ 
    \hline
\multicolumn{6}{r}{*$p$<0.05; **$p$<0.01; ***$p$<0.001 } \\
\end{tabular}
\end{center}
\end{table}

\subsection{Migrants tend to keep their X/Twitter Accounts.} 

The first experiment sought to determine whether users who migrated to alternative platforms were more or less likely to retain their X/Twitter accounts compared to those who did not migrate. Despite both the control and treatment groups initially comprising around 10,000 users, our data collection revealed that 8,347 users in the treatment group retained valid X/Twitter accounts, while 6,469 users from the control group still maintained their X/Twitter accounts. This discrepancy prompted our first research question: Do users migrating to alternative platforms exhibit different retention rates for their X/Twitter accounts?

Our initial hypothesis presumed that users migrating to alternative platforms would be less inclined to retain their X/Twitter accounts because platform migration might entail closing one account as another is opened. However, contrary to this hypothesis, our analysis, conducted with a $\chi^2$ test, revealed that those who migrated were more likely to retain their X/Twitter accounts than users who did not migrate ($\chi^2$(1, N = 13,280) = 160.16, $p$ < .001).

These unexpected findings raise questions about the underlying factors influencing user behavior during platform migration. One potential explanation could be the phenomenon of social media churn, wherein users who migrate between platforms tend to be more actively engaged on social media overall compared to those who do not migrate. However, without access to additional data, such as users' activity levels across various platforms, we cannot conclusively determine the underlying cause of this discrepancy.

\subsection{Highly Reputed X/Twitter accounts are more likely to Migrate.}

Next, our goal was to investigate (\textbf{RQ1}) whether highly reputed X/Twitter accounts were more likely to migrate to alternative platforms.

\begin{figure}[t]
    \centering
    \pgfplotstableread{
x          y    y-max  y-min
1  1.225 0.024   0.023
2  0.960 0.027   0.027
3     1.012 0.002   0.001
}{\mytable}

\begin{tikzpicture}
\begin{axis} [
    xticklabels={Followers,Following,Months},
    xtick={1,2,3},
    xlabel={},
    width=6cm,
    height=4cm,
    xmin=0.5,
    xmax=3.5,
    ylabel={Odds},
]
\addplot [only marks] 
  plot [error bars/.cd, y dir=both, y explicit]
  table [y error plus=y-max, y error minus=y-min] {\mytable};
\addplot[dashed, black] {1.00};
\end{axis} 
\end{tikzpicture}
    \caption{Odds plot (with 95\% confidence intervals) illustrating the probability that a user migrates to another platform for the number of followers, number of users following, and number of months spent on X/Twitter. Users are approximately 20\% more likely to migrate to another platform for each additionally follower (p<0.001).} 
    \label{fig:odds}
\end{figure}
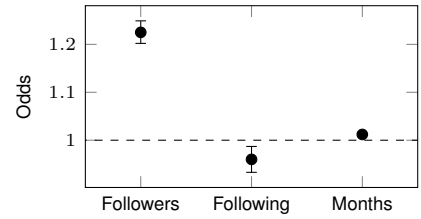

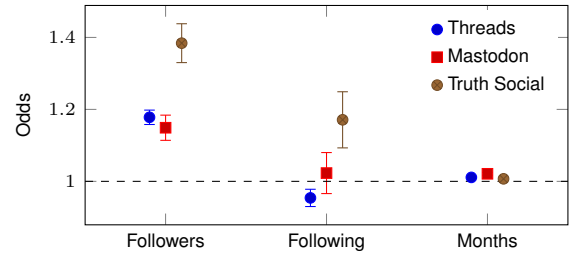
\begin{figure}[t]
    \centering
    \pgfplotsset{
  error bars/.cd,
    x dir=none,
    y dir=both, y explicit,
}

\begin{tikzpicture}
\begin{axis} [
    xticklabels={Followers, Following, Months},
    xtick={1,2,3},
    xlabel={},
    xmin=0.5,
    xmax=3.5,
    width=8cm,
    height=4.5cm,
    ylabel={Odds},
]
\addplot+ [only marks,] table [x expr={\thisrow{x}-0.1}, y=y, y error=ey]{
x          y    ey
1   1.178 0.020   
2   0.954 0.024   
3     1.011 0.001   
};

\addlegendentry{Threads}

\addplot+ [only marks,] table [x expr={\thisrow{x}}, y=y, y error=ey]{
x          y    ey  y-min
1  1.149 0.035   0.036
2  1.023 0.057   0.061
3     1.021 0.004   0.004
};

\addlegendentry{Mastodon}

\addplot+ [only marks,] table [x expr={\thisrow{x}+0.1}, y=y, y error=ey]{
x          y    ey  y-min
1  1.384 0.054   0.057
2  1.171 0.078   0.083
3     1.007 0.005   0.005
};

\addlegendentry{Truth Social}

\addplot[dashed, black,domain = 0.5:3.5] {1.00};

\end{axis} 
\end{tikzpicture}
    \caption{Odds plot (with 95\% confidence intervals) indicating that the number of followers is a strong predictor of migration, especially towards Truth Social.}
    \label{fig:breakdownoddsplot}
\end{figure}

The crawl of X/Twitter accounts enabled us to update the dataset with information on follower counts, account creation dates, and the number of accounts each user follows. We normalized the creation dates, considering dates prior to October 28, 2022, the day Elon Musk took control of Twitter, where positive values indicate accounts created before this event.

We used logistic regression to analyze the data, with the output variable representing whether the user migrated to another platform, encoded as 1 for migration and 0 for non-migration.

The results of the logistic regression, including logits and their corresponding t-values, are shown in Table~\ref{tab:logits}. Our analysis revealed significant correlations between migration and the number of followers, the number of accounts followed by each user, and their tenure on X/Twitter. However, interpreting the effects solely from logits can be challenging.

To provide a more intuitive understanding of the relationships, we illustrate the odds for each variable in Figure~\ref{fig:odds}. Overall, our findings indicate that a higher number of followers predicts migration, with each additional follower correlating with approximately a 20\% higher likelihood of migrating to an alternative platform. Moreover, longer tenure on X/Twitter slightly predicts migration, while the number of accounts followed by the user exhibits a slight inverse correlation with migration.

\begin{figure}[t]
    \centering
    \pgfplotsset{
  error bars/.cd,
    x dir=none,
    y dir=both, y explicit,
}

\begin{tikzpicture}
\begin{groupplot}[
    group style={
        group name=myplot,
        vertical sep=.2cm,
        horizontal sep=0.2cm,
        group size= 2 by 1
    },
    width=4.6cm,    
    height=3.5cm,
    xlabel={},
    x tick label style={rotate=45, anchor=east, font=\scriptsize},
    xticklabels={Threads, Mastodon, TruthSocial},
    xtick={1,2,3},
    ytick={0.1, 0, -0.1},
    legend cell align=left,
    y label style={at={(-0.2,0.5)}},
    title style={font=\sffamily\footnotesize, at={(0.5,0.9)}},
    ylabel={$\beta$},
    ymin=-0.11,
    ymax=0.11,
    legend image post style={scale=0.5},
    legend pos=north east,
    legend style={draw=none},
]
\nextgroupplot[title=Followers]

\addplot+ [only marks,] table [x expr={\thisrow{x}-0.1}, y=y, y error=ey]{
x          y    ey
1   0.079   0.001   
2   0.000   0.001  
3   0.0005  0.001  
};
\addplot[dashed, black,domain = 0.8:3.2] {0.00};

\nextgroupplot[title=Following, ylabel={}, yticklabels={}, title style={font=\sffamily\small, at={(0.5,0.86)}},]

\addplot+ [only marks,] table [x expr={\thisrow{x}-0.1}, y=y, y error=ey]{
x          y    ey
1   -0.072 0.091   
2   0.0008 0.001   
3   -0.0479 0.091   
};
\addplot[dashed, black,domain = 0.8:3.2] {0.00};
\end{groupplot}
\end{tikzpicture}
\begin{tikzpicture}
\begin{axis} [
    width=5.2cm,
    height=3.6cm,
    x tick label style={rotate=45, anchor=east, font=\scriptsize},
    xticklabels={Threads, Mastodon, TruthSocial},
    xtick={1,2,3},
    ymax = 110,
    ymin = -110,
    ytick={100, 0, -100},
    y label style={at={(-0.2,0.5)}},
    ylabel={$\beta$},
    xlabel={},
    title={Months},
    title style={font=\sffamily\footnotesize, at={(0.5,0.9)}},
]
\addplot+ [only marks,] table [x expr={\thisrow{x}-0.1}, y=y, y error=ey]{
x          y    ey
1   3.6631 29.607   
2   0.296 0.535   
3   51.532 42.962   
};
\addplot[dashed, black,domain = 0.8:3.2] {0.00};
\end{axis} 
\end{tikzpicture}







    \caption{Ordinary Least Squares (OLS) regression coefficient plots (and 95\% confidence intervals, sometimes covered by the mean-point) showing a statistical correlation in reputation (\textit{i.e.}, number of followers) between X/Twitter and Threads, but not between X/Twitter and other platforms.}
    \label{fig:reputationanalysis}
\end{figure}
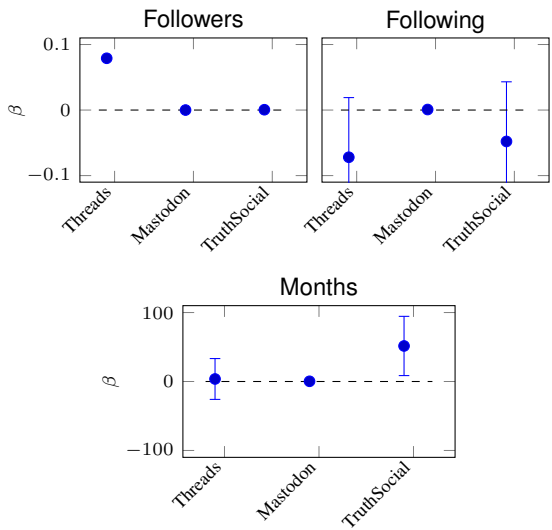

\subsection{Where did Users Migrate?}

Next, we performed a deeper analysis of the migration patterns by examining migration patterns to individual platforms: Threads, Mastodon, and Truth Social. Unlike the binary analysis in the previous section, which focused on whether users migrated or not, this analysis aimed to understand where users migrated to.

Recall that this set of usernames was contingent upon their presence on X/Twitter. Within this set, we identified 7,740 users on Threads, 1,019 users on Mastodon, and 380 accounts on Truth Social. Given these numbers, we ask: what is the relationship between users' activity on X/Twitter and their propensity to migrate to each respective platform?

We again employed logistic regression analysis to identify correlates for migration, considering variables such as the number of followers, the number of accounts the user follows, and their tenure on Twitter.

\begin{figure}[t]
    \centering
    \begin{tikzpicture}
\def\basis{1}
\pgfplotsset
{
y coord trafo/.code={\pgfmathparse{symlog(#1,\basis)}\pgfmathresult},
y coord inv trafo/.code={\pgfmathparse{symexp(#1,\basis)}\pgfmathresult},
}
\begin{axis}[
    width=4.6cm,    
    height=3.5cm,
    xtick={-1,0,1},
    ytick={-10000,0,10000},
    yticklabels={$-10^4$,$0$,$10^4$},
    ymajorgrids = true,
    xmajorgrids = true,      
    y label style={at={(-0.15,0.5)}},
    x label style={at={(0.5,-0.15)}},
    xlabel={e(intercept | $x$)}, ylabel={$\log$ e(followers|$x$)}, 
]
\addplot+ [only marks, mark size=0.75] table [x=x, y=y]{data.dat};
\addplot+ [only marks, mark size=0.75] table [x=x, y=y]{mastdata.dat};
\addplot+ [only marks, mark size=0.75] table [x=x, y=y]{TSdata.dat};
\end{axis}
\end{tikzpicture}
\begin{tikzpicture}
\def\basis{1}
\pgfplotsset
{
y coord trafo/.code={\pgfmathparse{symlog(#1,\basis)}\pgfmathresult},
y coord inv trafo/.code={\pgfmathparse{symexp(#1,\basis)}\pgfmathresult},
x coord trafo/.code={\pgfmathparse{symlog(#1,\basis)}\pgfmathresult},
x coord inv trafo/.code={\pgfmathparse{symexp(#1,\basis)}\pgfmathresult},
}
\begin{axis}[
    width=4.5cm,    
    height=3.5cm,   
    xtick={-10000,0,10000},
    ytick={-10000,0,10000},
    yticklabels={$-10^4$,$0$,$10^4$},
    xticklabels={$-10^4$,$0$,$10^4$},
    ymajorgrids = true,
    xmajorgrids = true,  
    y label style={at={(-0.15,0.5)}},
    x label style={at={(0.5,-0.15)}},
    xlabel={$\log$ e(followers | $x$)}, ylabel={$\log$ e(followers|$x$)}, 
]
\addplot+ [only marks, mark size=0.75] table [x=x, y=y]{data2.dat};
\addplot+ [only marks, mark size=0.75] table [x=x, y=y]{mastdata2.dat};
\addplot+ [only marks, mark size=0.75] table [x=x, y=y]{TSdata2.dat};

\end{axis}
\end{tikzpicture}
\begin{tikzpicture}
\def\basis{1}
\pgfplotsset
{
y coord trafo/.code={\pgfmathparse{symlog(#1,\basis)}\pgfmathresult},
y coord inv trafo/.code={\pgfmathparse{symexp(#1,\basis)}\pgfmathresult},
x coord trafo/.code={\pgfmathparse{symlog(#1,\basis)}\pgfmathresult},
x coord inv trafo/.code={\pgfmathparse{symexp(#1,\basis)}\pgfmathresult},
}
\begin{axis}[
    width=4.6cm,    
    height=3.5cm,
    xtick={-1000,0,1000},
    ytick={-10000,0,10000},
    ymajorgrids = true,
    xmajorgrids = true,    
    yticklabels={$-10^4$,$0$,$10^4$},
    xticklabels={$-10^4$,$0$,$10^4$},
    y label style={at={(-0.15,0.5)}},
    xlabel={$\log$ e(following | $x$)}, 
    ylabel={$\log$ e(followers|$x$)}, 
    x label style={at={(0.5,-0.15)}},
]
\addplot+ [only marks, mark size=0.75] table [x=x, y=y]{data3.dat};
\addplot+ [only marks, mark size=0.75] table [x=x, y=y]{mastdata3.dat};
\addplot+ [only marks, mark size=0.75] table [x=x, y=y]{TSdata3.dat};
\end{axis}
\end{tikzpicture}
\begin{tikzpicture}
\def\basis{1}
\pgfplotsset
{
y coord trafo/.code={\pgfmathparse{symlog(#1,\basis)}\pgfmathresult},
y coord inv trafo/.code={\pgfmathparse{symexp(#1,\basis)}\pgfmathresult},
}
\begin{axis}[
    width=4.5cm,    
    height=3.5cm,  
    xtick={-100,0,100},
    ytick={-10000,0,10000},
    yticklabels={$-10^4$,$0$,$10^4$},
    xticklabels={$-100$,$0$,$100$},
    ymajorgrids = true,
    xmajorgrids = true,
    y label style={at={(-0.15,0.5)}},
    x label style={at={(0.5,-0.15)}},
    xlabel={e(months | $x$)}, ylabel={$\log$ e(followers|$x$)},
]
\addplot+ [only marks, mark size=0.75] table [x=x, y=y]{data4.dat};
\addplot+ [only marks, mark size=0.75] table [x=x, y=y]{mastdata4.dat};
\addplot+ [only marks, mark size=0.75] table [x=x, y=y]{TSdata4.dat};
\end{axis}
\end{tikzpicture}
{\centering\sffamily\scriptsize
\begin{tabular}{lll}
    $\mathcolor{blue}{\bullet}$ Threads & \textcolor{red}{\rule{1mm}{1mm}} Mastodon & $\mathcolor{brown}{\bullet}$ Truth Social  \\
\end{tabular}
}
    \caption{Partial regression analysis illustrating the individual correlates for the number of followers on the alternative platforms as a function of number of followers, number of users following, and tenure (months) respectively. The number of followers from X/Twitter is significantly correlated with the number of followers on Threads (R=0.703, p<0.001), but not between X/Twitter and other platforms.}
    \label{fig:threadsregression}
\end{figure}
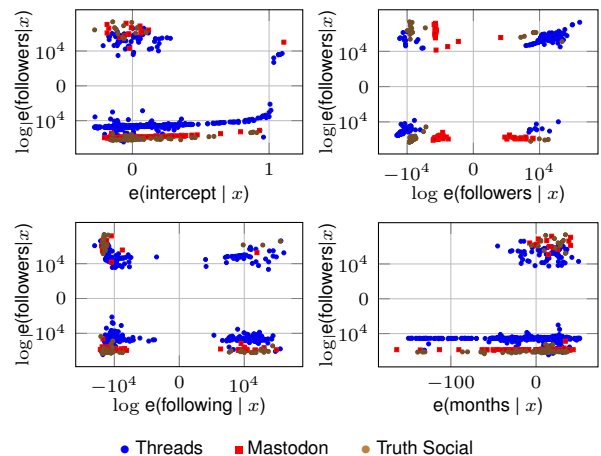

The odds for each variable for each platform are illustrated in Fig.\ref{fig:breakdownoddsplot}. Our findings indicate that the number of followers serves as a stronger predictor of migration to Truth Social compared to the other platforms. However, it is essential to note that this observation should be tempered by the relatively low number of users who migrated to Truth Social overall, as illustrated in Fig.\ref{fig:disapora}.

\section{Reputation Analysis}

The previous section focused solely on the likelihood of users migrating to alternative platforms. Next, we ask: (\textbf{RQ2}) When users do migrate, do their reputations, quantified by their number of followers, transfer with them?

To investigate this, we conducted Ordinary Least Squares (OLS) regression analyses for each group of users migrating to different platforms. Our hypothesis is that a strong correlation exists between the number of followers a user has on X/Twitter and the number of followers they accumulate on the platform to which they migrated.

Regression coefficients, along with their corresponding 95\% confidence intervals, are illustrated in Figure~\ref{fig:reputationanalysis}. Our analysis reveals a robust and statistically significant correlation in reputation between X/Twitter and Threads, but this reputation transfer was not found between X/Twitter and Mastodon nor between X/Twitter and Truth Social.

Finally, Figure~\ref{fig:threadsregression} illustrates partial regression plots for each independent variable against the dependent variable while holding the other variable constant. Each subplot in Fig.~\ref{fig:threadsregression} represents the correlation between reputation (number of followers on Threads, Mastodon, Truth Social) and one independent variable (such as the number of followers on X/Twitter or the number of months on X/Twitter) while controlling for the effects of other variables in the model.

We used Tukey HSD to perform a posthoc test on the pairwise comparisons. We find that reputation did transfers from X/Twitter to Threads; that is, the number of followers of a user on X/Twitter is significantly correlated with the number of followers on Threads (R=0.703, p<0.001). However, findings were not statistically significant for Mastodon and Truth Social (p=0.865 and p=0.912 resp.) nor for the number following and tenure variables.


\section{Conclusion}

The study reveals several key findings regarding user migration patterns, reputation transfer, and platform-specific differences in the context of social media dynamics: (1) users who migrate to alternative platforms were more likely to maintain their presence on X/Twitter compared to those who did not migrate; (2) the number of followers on X/Twitter positively correlated with the likelihood of migration to alternative platforms, indicating a relationship between user reputation and migration behavior; (3) users tended to preserve their reputation, measured by follower count, when transitioning from X/Twitter to Threads, but not to Mastodon or Truth Social. 

The implications of this study extend to various stakeholders in the social media landscape. First, platform developers and administrators can benefit from understanding the dynamics of user migration and reputation transfer, enabling them to tailor platform features and policies to better retain users and manage reputation transitions. Additionally, policymakers and regulatory bodies may find insights from this study valuable in shaping policies related to data portability, user privacy, and platform competition. For users, the study highlights the importance of reputation management and the potential impact of migration on their online presence and influence. 

Despite its contributions, this study has several limitations worth noting. The reliance on username matching as a method for identifying migrated users may introduce biases, as users may have different usernames across platforms or may not be captured due to variations in usernames. Second, this study compares the number of followers among usernames, not the actual sets of followers; although it is unlikely, it could be that the followers on one social media platform is disjoint from the set of followers on another. Third, the study focuses primarily on quantitative measures of reputation transfer, such as follower counts, overlooking qualitative aspects such as engagement and content quality. Additionally, the study's scope is limited to X/Twitter, Threads, Mastodon, and Truth Social, potentially overlooking trends and behaviors on other platforms. Finally, the study's analysis is based on observational data, limiting the ability to establish causal relationships between migration and reputation transfer. 

Future research in this area could explore several avenues to address the limitations of this study and expand our understanding of social media migration and reputation transfer. For example, qualitative studies could provide deeper insights into the motivations and experiences of users who migrate between platforms, shedding light on the underlying reasons for migration and the dynamics of reputation transfer. Furthermore, longitudinal studies could track users over time to examine how their reputations evolve after migration and whether any initial patterns persist or change over time.

In conclusion, this study sheds light on the complex dynamics of social media migration and reputation transfer, revealing insights into how users navigate between platforms and maintain their online identities. By analyzing user data from multiple platforms and employing statistical techniques, we have uncovered patterns of migration and identified factors influencing reputation transfer. While our findings provide valuable insights, there are still many unanswered questions and avenues for future research. Understanding these dynamics is crucial in an era where social media is changing dramatically and playing a large role in shaping public discourse.

\section*{Acknowledgements}
We would like to thank Maria Silvestri for her contributions to this paper. This work is supported by USAID \#7200AA18CA00059 and DARPA \#HR0011260595.



\end{document}